\documentclass[twocolumn,showpacs,preprintnumbers,amsmath,amssymb]{revtex4}

\usepackage{graphicx}
\usepackage{dcolumn}
\usepackage{bm}

\begin{document}


\title {A note on spin rescalings in post-Newtonian theory
} 
\author{Weiqun Jiang} 
\author{Xin Wu}
\email{xwu@ncu.edu.cn}
\affiliation{Department of Physics $\&$ Institute of Astronomy,
Nanchang University, Nanchang 330031, China}


\begin{abstract}

Usually, the reduced mass $\mu$ is viewed as a dropped factor in $\mu l$ and $\mu h$,
where $l$ and $h$ are dimensionless Lagrangian and Hamiltonian functions.
However, it must be retained in post-Newtonian systems of spinning compact binaries under a set of scaling spin transformations
$\mathbf{S}_i=\mathbf{s}_iGM^{2}$ because $l$ and $h$ do not keep the consistency of the orbital equations and the spin precession equations
but $(\mu/M) l$ and $(\mu/M) h$ do.
When another set of scaling spin transformations $\mathbf{S}_i=\mathbf{s}_iG\mu M$ are adopted,
the consistency of the orbital and spin equations is kept in $l$ or $h$, and the factor $\mu$ can be eliminated.
In addition, there are some other interesting results as follows.
The next-to-leading-order spin-orbit interaction is induced in the accelerations of
the simple Lagrangian of spinning compact binaries with the Newtonian and leading-order spin-orbit contributions,
and the next-to-leading-order spin-spin coupling is present in a post-Newtonian Hamiltonian that is exactly equivalent to the
Lagrangian formalism. If any truncations occur in the Euler-Lagrangian equations or the Hamiltonian,
then the Lagrangian and Hamiltonian formulations lose their equivalence. In fact,
the Lagrangian including the accelerations with or without truncations can be chaotic for the two bodies spinning,
whereas the Hamiltonian without the spin-spin term is integrable.

\end{abstract}


\maketitle

\section{Introduction}

Recently, the dynamics of spinning compact binaries has been a hot topic
since gravitational-wave signals from
two coalescing black holes or neutron stars were successfully, directly detected.
The post-Newtonian (PN) Lagrangian or Hamiltonian
formalism is often used for the description of this dynamics.
The physical equivalence of the two approaches at the same order was confirmed by
several authors [1-3]. However, a different claim on their nonequivalence was given in [4-6]
because higher-order PN terms are truncated during the transformation between them.

As in the study of usual  physical problems, dimensionless operations are widely implemented in the research of
spinning compact binaries so that the related variables, equations of motion and constants of motion become simple.
In the literature there are basically two sets of
dimensionless scaling spin transformations. One is $\mathbf{S}_i=\mathbf{s}_iG\mu M$, and the other is
$\mathbf{S}_i=\mathbf{s}_iGM^{2}$. The former spin transformations were used in many articles (e.g. [7-10])
because the dimensionless orbital angular momentum $\mathbf{r}\times \mathbf{\widetilde{p}}$ in the dimensionless
total angular momentum $\mathbf{\widetilde{j}} =\mathbf{r}\times \mathbf{\widetilde{p}}+(\mathbf{s}_1+\mathbf{s}_2)/c$
is measured in terms of $G\mu M$ and then the dimensionless spins $\mathbf{s}_1$ and $\mathbf{s}_2$
should also be measured in terms of the same unit system together.
The latter spin transformations considered in [4-6,11] are based on the fact that the magnitudes
of the spins $\mathbf{S}_i$ are $S_i=\chi_im^{2}_{i}$ ($i=1,2$) and $S_i/(GM^{2})=s_i$ become dimensionless.
Now, suppose that with the help of these dimensionless operations, a physical PN Lagrangian $L$ or Hamiltonian $H$  of spinning compact binaries
is readjusted as $L= \mu l$ or $H=\mu h$, where $l$ or $h$ is dimensionless.
A question is whether the factor $\mu$ in $\mu l$ and $\mu h$ can be dropped in the present case, as in the usual case.

To answer this question, we focus on the dimensionless operations of the simple Lagrangian formalism of spinning compact binaries
with the Newtonian and leading-order spin-orbit contributions.
The dimensionless operations of PN Hamiltonian formulations derived from the Lagrangian formalism are also considered.
It will be shown that this factor cannot be dropped for the spin transformations $\mathbf{S}_i=\mathbf{s}_iGM^{2}$ but can
for the spin transformations $\mathbf{S}_i=\mathbf{s}_iG\mu M$.
Thus, one of the main aims of this paper is to give a caution on the dimensionless operations
to systems of spinning compact binaries and to avoid the occurrence of errors
for the use of the spin transformations, such as $\mathbf{S}_i=\mathbf{s}_iGM^{2}$.
Another of the main aims is to provide an in-depth insight into the relationship between the PN Lagrangian and Hamiltonian formalisms.
Some other interesting results
are given through the transformation between the PN Lagrangian and Hamiltonian formulations.
The Euler-Lagrangian equations without truncations of higher-order terms have the next-to-leading-order spin-orbit interaction and
the Hamiltonian without any terms truncated  contains the next-to-leading-order spin-spin coupling.
The equivalence of the PN Lagrangian and Hamiltonian formulations are studied.
The integrability or nonintegrability and the chaoticity or regularity of these PN systems
are discussed.

This paper is organized as follows. In Sect. II we take a simple example of spinning compact binaries to investigate the dimensionless operations of
the PN Lagrangian and Hamiltonian formulations and the relationship between the PN approaches.
Using a generic PN system of spinning compact binaries,
we explain in Sect. III  why the factor $\mu$ is not eliminated under some scaling spin transformations.
Finally, the main results are concluded in Sect. IV.
At the end of this paper Appendix A relates to the spin precession equations that
can be derived from the Hamilton's canonical equations. Appendix B describes
the onset of chaos in the simple Lagrangian formalism including the accelerations with or without truncations.

In this paper we use vector formulas as follows: $\mathbf{a}\cdot(\mathbf{b}\times\mathbf{c})
=\mathbf{b}\cdot(\mathbf{c}\times\mathbf{a})=\mathbf{c}\cdot(\mathbf{a}\times\mathbf{b})$, $\mathbf{a}\times(\mathbf{b}\times\mathbf{c})
=(\mathbf{a}\cdot\mathbf{c})\mathbf{b}-(\mathbf{a}\cdot\mathbf{b})\mathbf{c}$, and $(\mathbf{a}\times\mathbf{b})\cdot(\mathbf{c}\times\mathbf{d})
=(\mathbf{a}\cdot\mathbf{c})(\mathbf{b}\cdot\mathbf{d})-(\mathbf{a}\cdot\mathbf{d})(\mathbf{b}\cdot\mathbf{c})$.

\section{A simple compact two-body system}

Consider a compact binary system that has masses $m_1$ and $m_2$,
the total mass $M=m_1+m_2$, the reduced mass
$\mu=m_1m_2/M$ and the mass ratio $\beta=m_1/m_2$.
The position and velocity of
body 1 relative to body 2 are $\mathbf{R}$ and $\mathbf{V}$, respectively.
$R=|\mathbf{R}|$ is the relative distance.
In addition, spin motions of the binaries are given by
$\mathbf{S}_1$ and $\mathbf{S}_2$, which have spin magnitudes $S_1=|\mathbf{S}_1|=\chi_1m^{2}_{1}$
and $S_2=|\mathbf{S}_2|=\chi_2m^{2}_{2}$ with dimensionless parameters $0\leq\chi_1\leq1$ and $0\leq\chi_2\leq1$.

\subsection{PN Lagrangian formulation}

For our purposes we take a simple form of the spinning compact two-body problem,
which contains the Newtonian and leading-order spin-orbit contributions in
the following Lagrangian formulation [12-14]
\begin{eqnarray}
\mathcal{L}= \frac{\mu}{2}\mathbf{V}^{2}+\frac{GM\mu}{R}+\frac{G\mu}{c^{3}R^{3}}\mathbf{V}\cdot
[\mathbf{R}\times (\gamma_1\mathbf{S}_1+\gamma_2\mathbf{S}_2)],
\end{eqnarray}
where $\gamma_1 = 2+3/(2\beta)$ and $\gamma_2 = 2+3\beta/2$.
The third term in Eq. (1) is the spin-orbit coupling with a 1.5 PN accuracy.
$G$ and $c$ are the gravitational constant and the speed of light, respectively.
Under the Newton-Wigner-Pryce spin supplementary condition, this Lagrangian
does not depend on accelerations [14]. However, it does
when other spin supplementary conditions
(such as the Frenkel-Mathisson-Pirani spin supplementary condition) are chosen in [13].

Define the generalized momenta
\begin{equation}
\mathbf{P}=\frac{\partial\mathcal{L}}{\partial\mathbf{V}}=\mu\mathbf{V}+\frac{G\mu}{c^{3}R^{3}}
\mathbf{R}\times (\gamma_1\mathbf{S}_1+\gamma_2\mathbf{S}_2).
\end{equation}
Based on the Euler-Lagrangian equation
\begin{equation}
\frac{d\mathbf{P}}{dT}=\frac{\partial\mathcal{L}}{\partial\mathbf{R}},
\end{equation}
the relative accelerations as the equations of motion can be derived.
Up to the 1.5PN order, they are
\begin{eqnarray}
\mathbf{A}_1 &=& -\frac{GM}{R^{3}}\mathbf{R}+\frac{G}{c^{3}R^{3}}\{\frac{3}{R^{2}}\mathbf{R}[(\mathbf{R}\times \mathbf{V})\cdot
(\gamma_1\mathbf{S}_1+\gamma_2\mathbf{S}_2)] \nonumber \\
&& -2\mathbf{V}\times(\gamma_1\mathbf{S}_1+\gamma_2\mathbf{S}_2)
\nonumber \\
&&  +\frac{3}{R^{2}}(\mathbf{R}\cdot \mathbf{V})[\mathbf{R}\times(\gamma_1\mathbf{S}_1+\gamma_2\mathbf{S}_2)]\}.
\end{eqnarray}
Up to a higher order, they are written as
\begin{equation}
\mathbf{A}_{2} =\mathbf{A}_{1}-\frac{G}{c^{3}R^{3}}\mathbf{R}\times(\gamma_1\mathbf{\dot{S}}_1+\gamma_2\mathbf{\dot{S}}_2).
\end{equation}
The spin precession equations about $\mathbf{\dot{S}}_1$ and $\mathbf{\dot{S}}_2$ can also be derived from the Euler-Lagrangian equation.
Their derivations are given as follows.
Setting $\mathbf{S}_1=\mathbf{B}_1\times \mathbf{C}_1$ and $\mathbf{S}_2=\mathbf{B}_2\times \mathbf{C}_2$,
we have the spin kinetic energy $(\mathbf{\dot{B}}_1\cdot \mathbf{C}_1-\mathbf{B}_1\cdot \mathbf{\dot{C}}_1)/2
+(\mathbf{\dot{B}}_2\cdot \mathbf{C}_2-\mathbf{B}_2\cdot \mathbf{\dot{C}}_2)/2$. When it is included in Eq. (1), the Lagrangian
becomes
\begin{eqnarray}
\mathcal{\overline{L}} &=& \frac{\mu}{2}\mathbf{V}^{2}+\frac{GM\mu}{R}+\frac{G\mu}{c^{3}R^{3}}\mathbf{V}\cdot
\{\mathbf{R}\times [\gamma_1(\mathbf{B}_1\times \mathbf{C}_1) \nonumber \\
& & +\gamma_2(\mathbf{B}_2\times \mathbf{C}_2)]\} +\frac{1}{2}(\mathbf{\dot{B}}_1\cdot \mathbf{C}_1-\mathbf{B}_1\cdot \mathbf{\dot{C}}_1) \nonumber \\
& & +\frac{1}{2}(\mathbf{\dot{B}}_2\cdot \mathbf{C}_2-\mathbf{B}_2\cdot \mathbf{\dot{C}}_2).
\end{eqnarray}
It is clear that Eqs. (4) and (5) can still be obtained from the Euler-Lagrangian equation of the modified Lagrangian $\mathcal{\overline{L}}$.
Additionally, the spins satisfy the Euler-Lagrangian equations
\begin{equation}
\frac{d}{dT}(\frac{\partial\mathcal{\overline{L}}}{\partial\mathbf{\dot{B}}_i})=\frac{\partial\mathcal{\overline{L}}}{\partial\mathbf{B}_i},
~~ \frac{d}{dT}(\frac{\partial\mathcal{\overline{L}}}{\partial\mathbf{\dot{C}}_i})=\frac{\partial\mathcal{\overline{L}}}{\partial\mathbf{C}_i},
~~ (i=1,2).
\end{equation}
They correspond to the following expressions
\begin{eqnarray}
\mathbf{\dot{B}}_i &=& \frac{G\mu\gamma_i}{c^{2}R^{3}}(\mathbf{R}\times \mathbf{V})\times \mathbf{B}_i, \\
\mathbf{\dot{C}}_i &=& \frac{G\mu\gamma_i}{c^{2}R^{3}}(\mathbf{R}\times \mathbf{V})\times \mathbf{C}_i.
\end{eqnarray}
Thus, we have the spin precession equations
\begin{eqnarray}
\mathbf{\dot{S}}_i &=&  \mathbf{\dot{B}}_i \times \mathbf{C}_i + \mathbf{B}_i \times \mathbf{\dot{C}}_i
=\frac{G\mu\gamma_i}{c^{2}R^{3}}(\mathbf{R}\times \mathbf{V})\times \mathbf{S}_i \nonumber \\
&=& \mathbf{S}_i\times\frac{\partial \mathcal{L}}{\partial \mathbf{S}_i}.
\end{eqnarray}
Eq. (5) is rewritten as
\begin{eqnarray}
\mathbf{A}_{2} &=& \mathbf{A}_{1}-\frac{G^{2}\mu}{c^{5}R^{6}}\{\mathbf{R}\times[(\mathbf{R}\times \mathbf{V})
 \nonumber \\ && \times(\gamma^{2}_1\mathbf{S}_1+\gamma^{2}_2\mathbf{S}_2)]\}.
\end{eqnarray}
It is worth noting that the leading-order spin-orbit contribution to the accelerations (4)
is at 1.5PN order, but to the spin precession equations (10) remains at 1PN order.
Note that the lowest order in the spin precession equations is the 1PN order (see Eq. (6.2) in [15]).
The second term in Eq. (11)
is a new term with a 2.5 PN accuracy. In fact, it belongs to a next-to-leading-order spin-orbit interaction.
The accelerations (11) without any terms truncated
consist of the Newtonian term and  the leading-order and next-to-leading-order spin-orbit interactions.

The system (1) has an energy
\begin{equation}
E =\mathbf{V}\cdot \mathbf{P}-\mathcal{L}= \frac{\mu}{2}\mathbf{V}^{2}-\frac{GM\mu}{R}.
\end{equation}
No matter whether the accelerations are expressed as Eq. (4) or Eq. (11), it is easy to check $dE/dT=0$,
that is, this energy is exactly  conserved.
There is also a strictly invariant total angular momentum vector
\begin{equation}
\mathbf{J} =\mathbf{R}\times \mathbf{P}+\frac{1}{c}\mathbf{S}_1+\frac{1}{c}\mathbf{S}_2.
\end{equation}
It should be pointed out that $\mu\mathbf{R}\times \mathbf{V}+(\mathbf{S}_1+\mathbf{S}_2)/c$ is not a constant of motion.
This is why the $\mathbf{P}$ in Eq. (13) uses the canonical total momentum (2) rather than the non-canonical
Newtonian momentum $\mu\mathbf{R}\times \mathbf{V}$.

\subsubsection{A set of scaling spin transformations: $\mathbf{S}_i=\mathbf{s}_iGM^{2}$}

For simplicity, dimensionless operations are often used. For the sake of this,
we adopt scale transformations to the position and time variables as follows:
\begin{equation}
\mathbf{R}=GM\mathbf{r}, ~~~~~~ T=GMt.
\end{equation}
In this case, we have
\begin{equation}
\mathbf{A}_{1}=\frac{1}{GM}\mathbf{a}_{1}, ~~~~~~ \mathbf{A}_{2}=\frac{1}{GM}\mathbf{a}_{2}.
\end{equation}
In view of the spin magnitudes $S_i=\chi_im^{2}_{i}$, it is reasonable to obtain dimensionless spin transformations
\begin{equation}
\mathbf{S}_i=\mathbf{s}_iGM^{2}.
\end{equation}
This can also be seen clearly from the spin-orbit terms without dimensions by
substituting Eqs. (14) and (15) into Eqs. (4) and (11). The scaling spin transformations
were adopted in references e.g. [4-6,11].
Therefore, the dimensionless accelerations are
\begin{eqnarray}
\mathbf{a}_1 &=& -\frac{\mathbf{r}}{r^{3}}+\frac{1}{c^{3}r^{3}}\{\frac{3}{r^{2}}\mathbf{r}[(\mathbf{r}\times \mathbf{v})\cdot
(\gamma_1\mathbf{s}_1+\gamma_2\mathbf{s}_2)] \nonumber \\
&& -2\mathbf{v}\times(\gamma_1\mathbf{s}_1+\gamma_2\mathbf{s}_2)
\nonumber \\
&&  +\frac{3}{r^{2}}(\mathbf{r}\cdot \mathbf{v})[\mathbf{r}\times(\gamma_1\mathbf{s}_1+\gamma_2\mathbf{s}_2)]\},
\end{eqnarray}
\begin{eqnarray}
\mathbf{a}_{2} &=& \mathbf{a}_{1}-\frac{1}{c^{5}r^{6}}\frac{\mu}{M}\{\mathbf{r}\times[(\mathbf{r}\times \mathbf{v})
 \nonumber \\ && \times(\gamma^{2}_1\mathbf{s}_1+\gamma^{2}_2\mathbf{s}_2)]\},
\end{eqnarray}
where $\mathbf{v}=d\mathbf{r}/dt$.
Noting Eqs. (14) and (16), we have
\begin{equation}
\mathbf{\dot{S}}_i=M\mathbf{\dot{s}}_i.
\end{equation}
In this sense, Eq. (10) is slightly modified as
\begin{eqnarray}
\mathbf{\dot{s}}_i=\frac{\gamma_i}{c^{2}r^{3}} \frac{\mu}{M}(\mathbf{r}\times \mathbf{v})\times \mathbf{s}_i.
\end{eqnarray}
Eq. (12) with $E=\mu\varepsilon$ is rewritten as
\begin{equation}
\varepsilon =\frac{1}{2}\mathbf{v}^{2}-\frac{1}{r}.
\end{equation}
Taking $\mathbf{J}=G\mu M \mathbf{j}$ and  $\mathbf{P}=\mu \mathbf{p}$ with $\mathbf{p}=\mathbf{v}+
\mathbf{r}\times (\gamma_1\mathbf{s}_1+\gamma_2\mathbf{s}_2)/(c^{3}r^{3})$, we modify Eq. (13) as
\begin{equation}
\mathbf{j} =\mathbf{r}\times \mathbf{p}+\frac{1}{c}\frac{M}{\mu}(\mathbf{s}_1+\mathbf{s}_2).
\end{equation}

The above dimensionless equations (17), (18) and (20) are obtained from the dimensionless operations to
the acceleration and spin equations (4), (10) and (11). What would happen if these dimensionless operations
are applied to the Lagrangian system (1)? To answer this question, we get $\mathcal{L}=\mu\mathcal{L}^{*}$
with the dimensionless Lagrangian
\begin{eqnarray}
\mathcal{L}^{*} &=& \frac{1}{2}\mathbf{v}^{2}+\frac{1}{r}+\mathcal{L}^{*}_{so},  \\
\mathcal{L}^{*}_{so} &=&
\frac{1}{c^{3}r^{3}}\mathbf{v}\cdot
[\mathbf{r}\times (\gamma_1\mathbf{s}_1+\gamma_2\mathbf{s}_2)]. \nonumber
\end{eqnarray}
The acceleration $\mathbf{a}_1$ and the energy $\varepsilon$
derived from the Lagrangian $\mathcal{L}^{*}$ are same as Eqs. (17) and (21), respectively.
However, the acceleration $\mathbf{a}^{*}_2$ and the spin equations $\mathbf{\dot{s}}^{*}_i$
do not contain  the factor $\mu/M$,
\begin{equation}
\mathbf{a}^{*}_{2} = \mathbf{a}_{1}-\frac{1}{c^{5}r^{6}}\{\mathbf{r}\times[(\mathbf{r}\times \mathbf{v}) \times(\gamma^{2}_1\mathbf{s}_1+\gamma^{2}_2\mathbf{s}_2)]\},
\end{equation}
\begin{equation}
\mathbf{\dot{s}}^{*}_i=\frac{\gamma_i}{c^{2}r^{3}} (\mathbf{r}\times \mathbf{v})\times \mathbf{s}_i.
\end{equation}
The dimensionless total angular momentum  $\mathbf{j}^{*}$ loses the factor $M/\mu$, either,
\begin{equation}
\mathbf{j}^{*} =\mathbf{r}\times \mathbf{p}+\frac{1}{c}(\mathbf{s}_1+\mathbf{s}_2).
\end{equation}
Which of the dimensionless expressions of $(\mathbf{a}_2,\mathbf{\dot{s}}_i,\mathbf{j})$ and
$(\mathbf{a}^{*}_2,\mathbf{\dot{s}}^{*}_i,\mathbf{j}^{*})$ is correct?
The former  expression is OK. Here are the related details on this answer.
The scale spin transformations (16) with Eq. (14) require that  $\mathbf{B}_i=\mathbf{b}_iM\sqrt{G}$, $\mathbf{C}_i=\mathbf{c}_iM\sqrt{G}$,
$\mathbf{\dot{B}}_i=\mathbf{\dot{b}}_i/\sqrt{G}$ and $\mathbf{\dot{C}}_i=\mathbf{\dot{c}}_i/\sqrt{G}$.
With the inclusion of dimensionless spin kinetic energy, we
have $\mathcal{\overline{L}}=\mu\mathcal{\overline{L}}^{*}$, where $\mathcal{\overline{L}}^{*}$ is a dimensionless Lagrangian
\begin{eqnarray}
\mathcal{\overline{L}}^{*} &=& \frac{1}{2}\mathbf{v}^{2}+\frac{1}{r}+\frac{1}{c^{3}r^{3}}\mathbf{v}\cdot
\{\mathbf{r}\times [\gamma_1(\mathbf{b}_1\times \mathbf{c}_1) \nonumber \\
& & +\gamma_2(\mathbf{b}_2\times \mathbf{c}_2)]\}  +\frac{1}{2}\frac{M}{\mu}(\mathbf{\dot{b}}_1\cdot
\mathbf{c}_1-\mathbf{b}_1\cdot \mathbf{\dot{c}}_1 \nonumber \\
& & +\mathbf{\dot{b}}_2\cdot \mathbf{c}_2-\mathbf{b}_2\cdot \mathbf{\dot{c}}_2).
\end{eqnarray}
When $\mathcal{\overline{L}}^{*}$ is used instead of $\mathcal{\overline{L}}$, Eqs. (8) and (9) become
\begin{eqnarray}
\mathbf{\dot{b}}_i &=& \frac{\mu}{M}\frac{\gamma_i}{c^{2}r^{3}}(\mathbf{r}\times \mathbf{v})\times \mathbf{b}_i, \\
\mathbf{\dot{c}}_i &=& \frac{\mu}{M}\frac{\gamma_i}{c^{2}r^{3}}(\mathbf{r}\times \mathbf{v})\times \mathbf{c}_i.
\end{eqnarray}
In this way, Eqs. (18) and (20) can be derived easily from $\mathcal{\overline{L}}^{*}$.
In fact, these results can also be obtained from $\mu\mathcal{L}^{*}$ rather than $\mathcal{L}^{*}$.
Note that the derivation of Eq. (20) is based on Eqs. (10) and (19), i.e. $\mathbf{\dot{S}}_i=M\mathbf{\dot{s}}_i=
\mathbf{s}_i\times [\partial (\mu\mathcal{L}^{*})/\partial \mathbf{s}_i]$.
Precisely speaking, Eqs. (17), (18) and (20) are completely determined by the same Lagrangian $(\mu/M)\mathcal{L}^{*}$.
That means that $\mu\mathcal{\overline{L}}^{*}$,  $\mathcal{\overline{L}}^{*}$
and $(\mu/M)\mathcal{L}^{*}$ are equivalent, i.e.  $\mu\mathcal{\overline{L}}^{*}\Leftrightarrow
\mathcal{\overline{L}}^{*}\Leftrightarrow (\mu/M)\mathcal{L}^{*}$,
whereas  $(\mu/M)\mathcal{L}^{*}$ and $\mathcal{L}^{*}$ are nonequivalent, i.e.  $(\mu/M)\mathcal{L}^{*}\nLeftrightarrow\mathcal{L}^{*}$.
In other words, the true expressions of the 1PN spin equations about $\mathbf{\dot{s}}_i$
rely on the modified  spin-orbit term
\begin{eqnarray}
\mathcal{\widetilde{L}}^{*}_{so}=\frac{\mu}{M}\mathcal{L}^{*}_{so} \nonumber
\end{eqnarray}
rather than the leading-order spin-orbit coupling $\mathcal{L}^{*}_{so}$.
This case is also suitable for
the expression of the 2.5PN term in the acceleration $\mathbf{a}_2$.
The factor $\mu$ in $\mu\mathcal{L}^{*}$  is not dropped.
Due to the dimensionless requirement, $\mu$ should give place to
$\mu/M$. In fact, $(\mu/M)\mathcal{L}^{*}$ is what we want.
Even if $\mu$ in $\mu\mathcal{L}^{*}$ is absent, $\mathbf{j}^{*}$ in Eq. (26) should be $\mathbf{j}$ in Eq. (22).

\subsubsection{Another set of scaling spin transformations: $\mathbf{S}_i=\mathbf{s}_iG\mu M$}

Since $\mathbf{R}\times \mathbf{P}=G\mu M\mathbf{r}\times \mathbf{p}$ during the transformation
$\mathbf{J}\rightarrow\mathbf{j}$ in Eqs. (13) and (22), $\mathbf{s}_i$ should also be measured in terms of the same unit system $G\mu M$,
\begin{equation}
\mathbf{S}_i=\mathbf{s}_iG\mu M.
\end{equation}
The scaling spin transformations were used in references such as [7-10].
In this case, the dimensionless accelerations corresponding to Eqs. (17) and (18) are
\begin{eqnarray}
\mathbf{\widetilde{a}}_1 &=& -\frac{\mathbf{r}}{r^{3}}+\frac{\mu}{M}\frac{1}{c^{3}r^{3}}\{\frac{3}{r^{2}}\mathbf{r}[(\mathbf{r}\times \mathbf{v})\cdot
(\gamma_1\mathbf{s}_1+\gamma_2\mathbf{s}_2)] \nonumber \\
&& -2\mathbf{v}\times(\gamma_1\mathbf{s}_1+\gamma_2\mathbf{s}_2)
\nonumber \\
&&  +\frac{3}{r^{2}}(\mathbf{r}\cdot \mathbf{v})[\mathbf{r}\times(\gamma_1\mathbf{s}_1+\gamma_2\mathbf{s}_2)]\},
\end{eqnarray}
\begin{eqnarray}
\mathbf{\widetilde{a}}_{2} &=& \mathbf{\widetilde{a}}_{1}-\frac{1}{c^{5}r^{6}}(\frac{\mu}{M})^{2}\{\mathbf{r}
\times[(\mathbf{r}\times \mathbf{v}) \nonumber \\ && \times(\gamma^{2}_1\mathbf{s}_1+\gamma^{2}_2\mathbf{s}_2)]\}.
\end{eqnarray}
The dimensionless spin equations are still Eq. (20). The acceleration and spin equations
can also given by the Lagrangian $\mu\mathcal{\widetilde{L}}^{*}$ or $\mathcal{\widetilde{L}}^{*}$
in the following form
\begin{eqnarray}
\mathcal{\widetilde{L}}^{*} = \frac{1}{2}\mathbf{v}^{2}+\frac{1}{r}+ \mathcal{\widetilde{L}}^{*}_{so}.
\end{eqnarray}
This fact shows the equivalence of $\mu\mathcal{\widetilde{L}}^{*}$ and $\mathcal{\widetilde{L}}^{*}$,
$\mu\mathcal{\widetilde{L}}^{*}\Leftrightarrow\mathcal{\widetilde{L}}^{*}$.
It is clear that $\mathcal{\widetilde{L}}^{*}$ in Eq. (33) is unlike $\mathcal{L}^{*}$ in Eq. (23) and determines
the true descriptions of the accelerations and the spin equations.
The energy does not depend on the spins and therefore is still Eq. (21).
However, the total angular momentum (22) depends on the spins and should be similar to Eq. (26), namely,
\begin{equation}
\mathbf{\widetilde{j}} =\mathbf{r}\times \mathbf{\widetilde{p}}+\frac{1}{c}(\mathbf{s}_1+\mathbf{s}_2),
\end{equation}
where $\mathbf{\widetilde{p}}=\mathbf{v}+
\mathbf{r}\times (\gamma_1\mathbf{s}_1+\gamma_2\mathbf{s}_2)\mu/(Mc^{3}r^{3})$.

\subsection{PN Hamiltonian formulations}

A Hamiltonian derived from the Lagrangian (1) is the same as the energy (12) in which
the velocity $\mathbf{V}$ must be expressed in terms of the canonical momentum  $\mathbf{P}$ (2),
\begin{equation}
\mathbf{V}=\frac{\mathbf{P}}{\mu}-\frac{G}{c^{3}R^{3}}
\mathbf{R}\times (\gamma_1\mathbf{S}_1+\gamma_2\mathbf{S}_2).
\end{equation}
The Hamiltonian up to the 1.5PN order is
\begin{equation}
H_1 =\frac{\mathbf{P}^{2}}{2\mu}-\frac{GM\mu}{R}+\frac{G}{c^{3} R^{3}}
(\mathbf{R}\times \mathbf{P})\cdot(\gamma_1\mathbf{S}_1+\gamma_2\mathbf{S}_2).
\end{equation}
The Hamiltonian up to a higher order is
\begin{eqnarray}
H_2 &=& H_1+\frac{\mu G^{2}}{2c^{6} R^{6}}
\{R^{2}(\gamma_1\mathbf{S}_1+\gamma_2\mathbf{S}_2)^{2}\nonumber \\
&& -[\mathbf{R}\cdot(\gamma_1\mathbf{S}_1 +\gamma_2\mathbf{S}_2)]^{2}\}.
\end{eqnarray}
The $o(c^{-6})$ term denotes a next-to-leading-order spin-spin interaction with a 3 PN accuracy.
It has appears in [4]. The energy $E$ in Eq. (12) is not exactly equal to but is approximately related to
$H_1$. In fact, they have a difference of the 3PN spin-spin term. Only $H_2$ is exactly identical to $E$, $E=H_2$.

The Hamilton's canonical equations for $H_1$ are
\begin{eqnarray}
\mathbf{\dot{R}}_{H_{1}} &=& \frac{\partial H_1}{\partial \mathbf{P}} =\frac{\mathbf{P}}{\mu}-\frac{G}{c^{3} R^{3}}
\mathbf{R}\times (\gamma_1\mathbf{S}_1+\gamma_2\mathbf{S}_2), \\
\mathbf{\dot{P}}_{H_{1}} &=& -\frac{\partial H_1}{\partial \mathbf{R}} =-\frac{GM\mu}{R^{3}}\mathbf{R}
+\frac{3G}{c^{3} R^{5}}\mathbf{R}
(\mathbf{R}\times \mathbf{P})\cdot(\gamma_1\mathbf{S}_1 \nonumber \\
&& +\gamma_2\mathbf{S}_2)
-\frac{G}{c^{3} R^{3}}
\mathbf{P}\times(\gamma_1\mathbf{S}_1+\gamma_2\mathbf{S}_2).
\end{eqnarray}
The spin equations for $H_1$ are given by
\begin{eqnarray}
\mathbf{\dot{S}}_{iH_{1}} = \frac{\partial H_1}{\partial \mathbf{S}_i}\times \mathbf{S}_i = \frac{G\gamma_i}{c^{2} R^{3}}
(\mathbf{R}\times \mathbf{P}) \times \mathbf{S}_i.
\end{eqnarray}
They are originated from the Hamilton's canonical equations, as shown in Appendix A.
For the Hamiltonian $H_2$, we also obtain the Hamilton's canonical equations
\begin{eqnarray}
\mathbf{\dot{R}}_{H_{2}} &=& \mathbf{\dot{R}}_{H_{1}}, \\
\mathbf{\dot{P}}_{H_{2}} &=& \mathbf{\dot{P}}_{H_{1}}+\frac{3\mu G^{2}}{c^{6} R^{8}}\mathbf{R}
\{R^{2}(\gamma_1\mathbf{S}_1+\gamma_2\mathbf{S}_2)^{2}-[\mathbf{R}\cdot(\gamma_1\mathbf{S}_1 \nonumber \\
&&  +\gamma_2\mathbf{S}_2)]^{2}\}
-\frac{\mu G^{2}}{c^{6} R^{6}}
\{\mathbf{R}(\gamma_1\mathbf{S}_1+\gamma_2\mathbf{S}_2)^{2}\nonumber \\
&& -[\mathbf{R}\cdot(\gamma_1\mathbf{S}_1 +\gamma_2\mathbf{S}_2)](\gamma_1\mathbf{S}_1 +\gamma_2\mathbf{S}_2)\},
\end{eqnarray}
\begin{eqnarray}
\mathbf{\dot{S}}_{iH_{2}} &=&\mathbf{\dot{S}}_{iH_{1}} +\frac{\gamma_i\mu G^{2}}{c^{5} R^{6}}
\{R^{2}(\gamma_1\mathbf{S}_1+\gamma_2\mathbf{S}_2)\nonumber \\
&& -[\mathbf{R}\cdot(\gamma_1\mathbf{S}_1 +\gamma_2\mathbf{S}_2)]\mathbf{R}\}\times \mathbf{S}_i.
\end{eqnarray}
Note that the second term in Eq. (43) has a 2.5 PN accuracy.

For the first spin transformations (16) with the above dimensionless treatments, Eqs. (38)-(43)
become
\begin{eqnarray}
\mathbf{\dot{r}}_{H_{1}} &=& \mathbf{p}-\frac{1}{c^{3} r^{3}}
\mathbf{r}\times (\gamma_1\mathbf{s}_1+\gamma_2\mathbf{s}_2), \\
\mathbf{\dot{p}}_{H_{1}} &=& -\frac{\mathbf{r}}{r^{3}}
+\frac{3}{c^{3} r^{5}}\mathbf{r}
[(\mathbf{r}\times \mathbf{p})\cdot(\gamma_1\mathbf{s}_1 +\gamma_2\mathbf{s}_2)] \nonumber \\
&&
-\frac{1}{c^{3} r^{3}}
\mathbf{p}\times(\gamma_1\mathbf{s}_1+\gamma_2\mathbf{s}_2),
\end{eqnarray}
\begin{eqnarray}
\mathbf{\dot{s}}_{iH_{1}} =  \frac{\mu\gamma_i}{c^{2}M r^{3}}
(\mathbf{r}\times \mathbf{p}) \times \mathbf{s}_i;
\end{eqnarray}
\begin{eqnarray}
\mathbf{\dot{r}}_{H_{2}} &=& \mathbf{\dot{r}}_{H_{1}}, \\
\mathbf{\dot{p}}_{H_{2}} &=& \mathbf{\dot{p}}_{H_{1}}+\frac{3}{c^{6} r^{8}}\mathbf{r}
\{r^{2}(\gamma_1\mathbf{s}_1+\gamma_2\mathbf{s}_2)^{2} \nonumber \\
&&  -[\mathbf{r}\cdot(\gamma_1\mathbf{s}_1+\gamma_2\mathbf{s}_2)]^{2}\}
-\frac{1}{c^{6} r^{6}}
\{\mathbf{r}(\gamma_1\mathbf{s}_1+\gamma_2\mathbf{s}_2)^{2}\nonumber \\
&& -[\mathbf{r}\cdot(\gamma_1\mathbf{s}_1 +\gamma_2\mathbf{s}_2)](\gamma_1\mathbf{s}_1 +\gamma_2\mathbf{s}_2)\},
\end{eqnarray}
\begin{eqnarray}
\mathbf{\dot{s}}_{iH_{2}} &=&\mathbf{\dot{s}}_{iH_{1}} +\frac{\gamma_i\mu}{c^{5}M r^{6}}
\{r^{2}(\gamma_1\mathbf{s}_1+\gamma_2\mathbf{s}_2)\nonumber \\
&& -[\mathbf{r}\cdot(\gamma_1\mathbf{s}_1 +\gamma_2\mathbf{s}_2)]\mathbf{r}\}\times \mathbf{s}_i.
\end{eqnarray}
Most of them can also be derived directly from the dimensionless Hamiltonians:
\begin{equation}
H^{*}_1 =\frac{\mathbf{p}^{2}}{2}-\frac{1}{r}+\frac{1}{c^{3} r^{3}}
(\mathbf{r}\times \mathbf{p})\cdot(\gamma_1\mathbf{s}_1+\gamma_2\mathbf{s}_2),
\end{equation}
\begin{eqnarray}
H^{*}_2 &=& H^{*}_1+\frac{1}{2c^{6} r^{6}}
\{r^{2}(\gamma_1\mathbf{s}_1+\gamma_2\mathbf{s}_2)^{2}\nonumber \\
&& -[\mathbf{r}\cdot(\gamma_1\mathbf{s}_1 +\gamma_2\mathbf{s}_2)]^{2}\}.
\end{eqnarray}
Obviously, Eqs. (44), (45), (47) and (48) can be given by $H^{*}_{1}$ and $H^{*}_{2}$, whereas Eqs. (46) and (49) can not.
In fact, Eqs. (46) and (49) are obtained from  $d\mathbf{S}_{iH_{1}}/dT=M\mathbf{\dot{s}}_{iH_{1}}
=\mu(\partial H^{*}_{1}/\partial \mathbf{s}_i)\times \mathbf{s}_i$ and $d\mathbf{S}_{iH_{2}}/dT=M\mathbf{\dot{s}}_{iH_{2}}
=\mu(\partial H^{*}_{2}/\partial \mathbf{s}_i)\times \mathbf{s}_i$.
Thus, the orbital equations (44), (45), (47) and (48) and the spin equations $\mathbf{\dot{s}}_{i}$ (46) and (49)
are not determined by the same Hamiltonians; the former comes
from $H^{*}_{1}$ and $H^{*}_{2}$, but the latter is derived from  $(\mu/M)H^{*}_{1}$ and $(\mu/M)H^{*}_{2}$.
Now, we can say that $\mu$ in $\mu H^{*}_{1}$ and $\mu H^{*}_{2}$ must be retained.
In our later discussions, $\mu$ in $\mu H^{*}_{1}$ and $\mu H^{*}_{2}$ should be replaced with $\mu/M$.
This is what the dimensionless operations require.

On the other hand, when the second spin transformations (30) are considered, Eqs. (38), (39), (42), (43)
are readjusted as
\begin{eqnarray}
\mathbf{\dot{r}}_{H_{1}} &=& \mathbf{p}-\frac{\mu}{M}\frac{1}{c^{3} r^{3}}
\mathbf{r}\times (\gamma_1\mathbf{s}_1+\gamma_2\mathbf{s}_2), \\
\mathbf{\dot{p}}_{H_{1}} &=& -\frac{\mathbf{r}}{r^{3}}
+\frac{\mu}{M}\frac{3}{c^{3} r^{5}}\mathbf{r}
[(\mathbf{r}\times \mathbf{p})\cdot(\gamma_1\mathbf{s}_1 +\gamma_2\mathbf{s}_2)] \nonumber \\
&&
-\frac{\mu}{M}\frac{1}{c^{3} r^{3}}
\mathbf{p}\times(\gamma_1\mathbf{s}_1+\gamma_2\mathbf{s}_2),
\end{eqnarray}
\begin{eqnarray}
\mathbf{\dot{p}}_{H_{2}} &=& \mathbf{\dot{p}}_{H_{1}}+(\frac{\mu}{M})^{2}\frac{3}{c^{6} r^{8}}\mathbf{r}
\{r^{2}(\gamma_1\mathbf{s}_1+\gamma_2\mathbf{s}_2)^{2} \nonumber \\
&&  -[\mathbf{r}\cdot(\gamma_1\mathbf{s}_1+\gamma_2\mathbf{s}_2)]^{2}\}
\nonumber \\
&& -(\frac{\mu}{M})^{2}\frac{1}{c^{6} r^{6}}\{\mathbf{r}(\gamma_1\mathbf{s}_1+\gamma_2\mathbf{s}_2)^{2} \nonumber \\
&& -[\mathbf{r}\cdot(\gamma_1\mathbf{s}_1 +\gamma_2\mathbf{s}_2)](\gamma_1\mathbf{s}_1 +\gamma_2\mathbf{s}_2)\},
\end{eqnarray}
\begin{eqnarray}
\mathbf{\dot{s}}_{iH_{2}} &=&\mathbf{\dot{s}}_{iH_{1}} +(\frac{\mu}{M})^{2}\frac{\gamma_i}{c^{5} r^{6}}
\{r^{2}(\gamma_1\mathbf{s}_1+\gamma_2\mathbf{s}_2)\nonumber \\
&& -[\mathbf{r}\cdot(\gamma_1\mathbf{s}_1 +\gamma_2\mathbf{s}_2)]\mathbf{r}\}\times \mathbf{s}_i.
\end{eqnarray}
The 1PN spin equations are Eq. (46). The canonical equations of motion and the spin equations
are also determined by the dimensionless Hamiltonians
\begin{equation}
\widetilde{H}^{*}_1 =\frac{\mathbf{p}^{2}}{2}-\frac{1}{r}+\frac{\mu}{M}\frac{1}{c^{3} r^{3}}
(\mathbf{r}\times \mathbf{p})\cdot(\gamma_1\mathbf{s}_1+\gamma_2\mathbf{s}_2),
\end{equation}
\begin{eqnarray}
\widetilde{H}^{*}_2 &=& H^{*}_1+(\frac{\mu}{M})^{2}\frac{1}{2c^{6} r^{6}}
\{r^{2}(\gamma_1\mathbf{s}_1+\gamma_2\mathbf{s}_2)^{2}\nonumber \\
&& -[\mathbf{r}\cdot(\gamma_1\mathbf{s}_1 +\gamma_2\mathbf{s}_2)]^{2}\}.
\end{eqnarray}
Of course, $\mu\widetilde{H}^{*}_1$ and $\widetilde{H}^{*}_1$ provide the same  canonical equations and spin equations,
i.e. $\mu\widetilde{H}^{*}_1\Leftrightarrow\widetilde{H}^{*}_1$. In addition, we have $\mu\widetilde{H}^{*}_2\Leftrightarrow\widetilde{H}^{*}_2$.

Here are several remarks about the aforementioned transformations from the dimensionless PN Lagrangian formulations to
the dimensionless PN Hamiltonian formulations.

\textbf{Remark 1}: The Hamiltonian quantity that is exactly
equal to the conserved energy $\varepsilon$ (21) should be $H^{*}_2$ (51) or $\widetilde{H}^{*}_2$ (57)
rather than $H^{*}_1$ (50) or $\widetilde{H}^{*}_1$ (56). In fact,
the 3PN spin-spin term is the difference between $\varepsilon$ and $H^{*}_1$ (or $\widetilde{H}^{*}_1$).
In this sense, $H^{*}_1$ or $\widetilde{H}^{*}_1$  is only approximately related to  $\varepsilon$.

\textbf{Remark 2}: As far as the equivalence between the Lagrangians or the Hamiltonians is concerned,
$\mu\mathcal{\overline{L}}^{*}\Leftrightarrow\mathcal{\overline{L}}^{*}\Leftrightarrow (\mu/M)\mathcal{L}^{*}\Leftrightarrow
\mu\mathcal{\widetilde{L}}^{*} \Leftrightarrow \mathcal{\widetilde{L}}^{*}$, $(\mu/M) H^{*}_{1}\Leftrightarrow\mu \widetilde{H}^{*}_{1}\Leftrightarrow\widetilde{H}^{*}_{1}$,
and $(\mu/M) H^{*}_{2}\Leftrightarrow\mu \widetilde{H}^{*}_{2}\Leftrightarrow\widetilde{H}^{*}_{2}$.
However, $(\mu/M)\mathcal{L}^{*}\nLeftrightarrow\mathcal{L}^{*}$, $(\mu/M) H^{*}_{1}\nLeftrightarrow H^{*}_{1}$,
and $(\mu/M) H^{*}_{2}\nLeftrightarrow H^{*}_{2}$. In other words, the factor $\mu$ or $\mu/M$ in
$(\mu/M)\mathcal{L}^{*}$, $(\mu/M) H^{*}_{1}$ and $(\mu/M) H^{*}_{2}$ cannot be dropped but can in $\mu\mathcal{\overline{L}}^{*}$
for the spin transformations (16). Since  $\mu$ or $\mu/M$ does not appear as a factor of $\mathcal{L}^{*}$, $H^{*}_{1}$ and $H^{*}_{2}$
under the spin transformations adopted in [4-6,11],
the dimensionless spin equations given in these references lose the factor $\mu/M$ and should have minor errors.
For the spin transformations (30),
the factor $\mu$ can be eliminated in $\mu\mathcal{\widetilde{L}}^{*}$, $\mu \widetilde{H}^{*}_{1}$
and $\mu \widetilde{H}^{*}_{2}$ without question.

\textbf{Remark 3}: As far as the equivalence between the PN Lagrangian and Hamiltonian formulations is concerned,
$(\mu/M) H^{*}_{2}$ $\Leftrightarrow$ $\mathcal{\overline{L}}^{*}$ with $\mathbf{a}_2$,
and $\widetilde{H}^{*}_{2}$ $\Leftrightarrow$ $\mathcal{\widetilde{L}}^{*}$ with $\mathbf{\widetilde{a}}_2$.
Only when no truncations of higher-order PN terms occur during the transformation between the Lagrangian and Hamiltonian formulations,
does this equivalence of $(\mu/M) H^{*}_{2}$ and $\mathcal{\overline{L}}^{*}$ with $\mathbf{a}_2$ (or $\widetilde{H}^{*}_{2}$ and $\mathcal{\widetilde{L}}^{*}$ with $\mathbf{\widetilde{a}}_2$) exist.
However, $(\mu/M) H^{*}_{1}$ $\nLeftrightarrow$ $(\mu/M) H^{*}_{2}$ $\nLeftrightarrow$ $\mathcal{\overline{L}}^{*}$ with $\mathbf{a}_1$,
and $\widetilde{H}^{*}_{1}$ $\nLeftrightarrow$ $\widetilde{H}^{*}_{2}$
$\nLeftrightarrow$ $\mathcal{\widetilde{L}}^{*}$ with $\mathbf{\widetilde{a}}_1$.
This is because some higher-order PN terms are always truncated during the transformations between them.
Thus, the PN Lagrangian and Hamiltonian formulations at the same order are nonequivalent in general.
These facts support the work [4] again.

\textbf{Remark 4}: Discussions on integrability and nonintegrability. The Hamiltonian $\widetilde{H}^{*}_{1}$
has four integrals of motion, involving the conserved Hamiltonian quantity (56) and the constant total angular momentum vector
(34). There is a fifth integral of motion, the conserved length of the Newtonian-like  angular momentum $\mathbf{r}\times \mathbf{p}$.
Although the spin magnitudes determined by Eq. (46) remain invariant, they are not viewed as
some of the five integrals of motion. In fact, they play an important role in constructing the canonical, conjugate spin variables [16].
When the two bodies spin, this Hamiltonian contains the five integrals in a ten-dimensional phase space quipped with a complete symplectic structure.
Based on Liouville's theorem about the  integrability of a canonical Hamiltonian system, this system should be integrable and nonchaotic.
On the other hand, when the 3PN spin-spin term is included in $\widetilde{H}^{*}_{2}$, the four integrals (21) and (34) are still present
but the length of the Newtonian-like angular momentum is no longer a constant. Therefore, $\widetilde{H}^{*}_{2}$ is nonintegrable and
can be chaotic in appropriate conditions. Of course, this result is also suitable for its equivalent Lagrangian formulation $\mathcal{\widetilde{L}}^{*}$ with $\mathbf{\widetilde{a}}_2$
[i.e. Eqs. (20) and (32)]. Even the Lagrangian formulation $\mathcal{\widetilde{L}}^{*}$ with $\mathbf{\widetilde{a}}_1$
[i.e. Eqs. (20) and (31)] is nonintegrable although we have no way to give its equivalent Hamiltonian and to prove
its nonintegrability in the analytical method. It is easy to check the onset of chaos in each of the two approaches $\mathcal{\widetilde{L}}^{*}$ with $\mathbf{\widetilde{a}}_2$ and $\mathcal{\widetilde{L}}^{*}$ with $\mathbf{\widetilde{a}}_1$ via a numerical technique in Appendix B.
It is worth emphasizing again that the Lagrangian formulation $\mathcal{\widetilde{L}}^{*}$ with $\mathbf{\widetilde{a}}_1$
and the Hamiltonian formulation $\widetilde{H}^{*}_{1}$ exhibit completely distinct dynamical behaviors
although only the Newtonian and leading-order spin-orbit contributions are considered in the two formulations.

\section{Explanations}

Why cannot  the factor $\mu$ in
$\mu\mathcal{L}^{*}$, $\mu H^{*}_{1}$ and $\mu H^{*}_{2}$ be dropped for the spin transformations (16)?
Why can it be eliminated in $\mu\mathcal{\overline{L}}^{*}$
for the spin transformations (16) and in $\mu\mathcal{\widetilde{L}}^{*}$, $\mu \widetilde{H}^{*}_{1}$
and $\mu \widetilde{H}^{*}_{2}$ for the spin transformations (30)?
To answer these questions, we take into account a
generic PN system of spinning compact binaries $L(\mathbf{R},\mathbf{V},\mathbf{S}_1,\mathbf{S}_2)$
or $H(\mathbf{R},\mathbf{P},\mathbf{S}_1,\mathbf{S}_2)$. With the aid of the aforementioned scale transformations,
$L$ and $H$ are readjusted as
\begin{eqnarray}
L(\mathbf{R},\mathbf{V},\mathbf{S}_1,\mathbf{S}_2) &=& \mu l(\mathbf{r},\mathbf{v},\mathbf{s}_1,\mathbf{s}_2), \\
H(\mathbf{R},\mathbf{P},\mathbf{S}_1,\mathbf{S}_2) &=& \mu h(\mathbf{r},\mathbf{p},\mathbf{s}_1,\mathbf{s}_2),
\end{eqnarray}
where $l$ and $h$ are dimensionless.

When the spin transformations (16) are adopted, the Euler-Lagrangian equation for $L$ is
\begin{equation}
\frac{d}{d(GMt)}(\frac{\partial (\mu l)}{\partial \mathbf{v}})=\frac{\partial(\mu l)}{\partial (GM\mathbf{r})}.
\end{equation}
The two factors $\mu$ in the left and right sides of the equality can be omitted without doubt. Thus, we have
\begin{equation}
\frac{d}{dt}(\frac{\partial l}{\partial \mathbf{v}})=\frac{\partial l}{\partial \mathbf{r}}.
\end{equation}
However, the factor $\mu$ appears only in the right side of the spin equations
\begin{equation}
\frac{d (GM^{2}\mathbf{s}_i)}{d(GMt)} =\mathbf{s}_i\times\frac{\partial(\mu l)}{\partial \mathbf{s}_i}.
\end{equation}
In this case, this factor is not eliminated. Now, the equations
are simplified as
\begin{equation}
\frac{d \mathbf{s}_i}{dt} =\frac{\mu}{M}\mathbf{s}_i\times\frac{\partial l}{\partial \mathbf{s}_i}
=\mathbf{s}_i\times\frac{\partial }{\partial \mathbf{s}_i}(\frac{\mu}{M}l).
\end{equation}
Seen from Eqs. (61) and (63), the Euler-Lagrangian equations and the spin equations seem to be from the two different
Lagrangian formalisms $l$ and $(\mu/M)l$. In fact, they are given by the same Lagrangian $(\mu/M)l$.
This is what we have shown in Eqs. (17), (18) and (20).
Therefore, the factor $\mu$ in $\mu\mathcal{L}^{*}$ or $\mu/M$ in $(\mu/M)\mathcal{L}^{*}$ must be retained.
Unlike in $\mu\mathcal{L}^{*}$, $\mu$ in $\mu\mathcal{\overline{L}}^{*}$ can be dropped
because it exists in the two sides of the spin equation (10) with $\mathcal{L}=\mu\mathcal{\overline{L}}^{*}$.
On the other hand, the Hamilton's canonical equations for $H$ are
\begin{eqnarray}
\frac{d (GM\mathbf{r})}{d(GMt)} = \frac{\partial (\mu h)}{\partial (\mu\mathbf{p})},  ~~~
\frac{d (\mu\mathbf{p})}{d(GMt)} = -\frac{\partial (\mu h)}{\partial (GM\mathbf{r})},
\end{eqnarray}
which are expressed as
\begin{eqnarray}
\frac{d \mathbf{r}}{dt} = \frac{\partial h}{\partial \mathbf{p}},  ~~~~
\frac{d \mathbf{p}}{dt} = -\frac{\partial h}{\partial \mathbf{r}}.
\end{eqnarray}
The spin equations for $H$ are
\begin{equation}
\frac{d (GM^{2}\mathbf{s}_i)}{d(GMt)} =\frac{\partial(\mu h)}{\partial \mathbf{s}_i}\times \mathbf{s}_i,
\end{equation}
which are written in the following form
\begin{equation}
\frac{d \mathbf{s}_i}{dt} =\frac{\mu}{M}\frac{\partial h}{\partial \mathbf{s}_i}\times \mathbf{s}_i.
\end{equation}
It is clear that the orbital equations (65) are given by the Hamiltonian $h$, but the spin equations (67) are obtained from
another Hamiltonian $(\mu/M) h$. Therefore, $\mu$ in $\mu h$ is not eliminated. The dimensionless operations
require that $\mu$ in $\mu h$ should be replaced with $\mu/M$. However, Eq. (65) is not suitable for $(\mu/M) h$.
To solve this question, we give scale transformations to the dimensionless coordinate $\mathbf{r}$ and the dimensionless momentum $\mathbf{p}$,
\begin{equation}
\mathbf{r}=\mathbf{\hat{r}}\sqrt{\frac{M}{\mu}}, ~~~~ \mathbf{p}=\mathbf{\hat{p}}\sqrt{\frac{M}{\mu}}.
\end{equation}
In fact, $\mathbf{\hat{r}}$ and $\mathbf{\hat{p}}$ are measured in terms of $GM\sqrt{M/\mu}$ and $\sqrt{\mu M}$, respectively.
Therefore, the Hamiltonian $(\mu/M) h$ can determine not only the spin equations (67) but also the orbital equations
\begin{eqnarray}
\frac{d \mathbf{\hat{r}}}{dt} = \frac{\mu}{M}\frac{\partial h}{\partial \mathbf{\hat{p}}},  ~~~~
\frac{d \mathbf{\hat{p}}}{dt} = -\frac{\mu}{M}\frac{\partial h}{\partial \mathbf{\hat{r}}}.
\end{eqnarray}
The dimensionless total angular momentum (22) is readjusted as
\begin{equation}
\mathbf{\hat{j}} =\mathbf{\hat{r}}\times \mathbf{\hat{p}}+\frac{1}{c}(\mathbf{s}_1+\mathbf{s}_2),
\end{equation}
which is measured in terms of $GM^{2}$. Although $\mathbf{\hat{j}}$ and $\mathbf{\widetilde{j}}$ in Eq. (34)
have the same expression, $\mathbf{\widetilde{j}}$ is measured in terms of $GM\mu$.
These demonstrations display that $\mu$ in
$\mu H^{*}_{1}$ and $\mu H^{*}_{2}$ or $\mu/M$ in
$(\mu/M) H^{*}_{1}$ and $(\mu/M) H^{*}_{2}$ is not dropped.

When the spin transformations (30) are used, there are still Eqs. (61) and (65).
The spin equations  for the Lagrangian $L$ are
\begin{equation}
\frac{d (GM\mu\mathbf{s}_i)}{d(GMt)} =\mathbf{s}_i\times\frac{\partial(\mu l)}{\partial \mathbf{s}_i},
\end{equation}
which become of the form
\begin{equation}
\frac{d \mathbf{s}_i}{dt} =\mathbf{s}_i\times\frac{\partial l}{\partial \mathbf{s}_i}.
\end{equation}
The spin equations  for the Hamiltonian $H$ are
\begin{equation}
\frac{d (GM\mu\mathbf{s}_i)}{d(GMt)} =\frac{\partial(\mu h)}{\partial \mathbf{s}_i}\times \mathbf{s}_i,
\end{equation}
which are expressed as
\begin{equation}
\frac{d \mathbf{s}_i}{dt} =\frac{\partial h}{\partial \mathbf{s}_i}\times \mathbf{s}_i.
\end{equation}
Under the spin transformations (30), the orbital equations and the spin equations use the same
Lagrangian or Hamiltonian formalism.
These facts show that $\mu$ in $\mu\mathcal{\widetilde{L}}^{*}$, $\mu \widetilde{H}^{*}_{1}$
and $\mu \widetilde{H}^{*}_{2}$ can be  omitted.

\section{Summary}

 In this paper, we mainly discuss dimensionless operations of the PN Lagrangian and Hamiltonian formulations
 of spinning compact binaries, $L= \mu l$ and
$H=\mu h$, where $l$ and $h$ are dimensionless.  For the spin transformations $\mathbf{S}_i=\mathbf{s}_iGM^{2}$,
 the orbital and spin precession equations are obtained from the same dimensionless Lagrangian formalism $(\mu/M)l$
or the same dimensionless Hamiltonian $(\mu/M)h$.
In this case, $\mu$  in $\mu l$  and
$\mu h$ or $\mu/M$ in $(\mu/M)l$ and $(\mu/M) h$ cannot be dropped. Because the factor $\mu$ or $\mu/M$ in the dimensionless PN Lagrangian and Hamiltonian formulations
is missing under the scaling spin transformations chosen in [4-6,11], the dimensionless spin equations in these references
lose the factor $\mu/M$ and have minor errors. Although these errors do not exert any influence on the main results
of the articles, minor corrections are still necessary and the factor $\mu/M$
should be added to the right functions of the dimensionless spin differential equations.
On the other hand,
both the Lagrangian $l$ and the Hamiltonian $h$
keep the consistency of the orbital and spin equations for the spin transformations $\mathbf{S}_i=\mathbf{s}_iG\mu M$.
Clearly, $\mu$ in $\mu l$ and $\mu h$ can be dropped.
Considering these facts, one should be cautious to carry out dimensionless operations of these problems
when different scaling spin transformations are employed.

Apart from the above point that should be cautioned during these dimensionless operations,
some other interesting results can be seen clearly via the transformation from the simple Lagrangian of spinning compact binaries
including the Newtonian and leading-order spin-orbit terms to the PN Hamiltonian formulations.
To our surprise, the next-to-leading-order spin-orbit term appears in the accelerations from the Euler-Lagrangian equations,
and the next-to-leading-order spin-spin coupling exists in the PN Hamiltonian formulation without any truncations.
The Lagrangian with the accelerations up to the 2.5PN order is exactly equivalent to the PN Hamiltonian up to the 3PN order,
but the Lagrangian with the accelerations up to the 1.5PN order is not equivalent to the Hamiltonian up to the 1.5PN or 3PN order.
For the two bodies spinning, the 1.5PN Hamiltonian is integrable due to the presence of five integrals,
whereas the 3PN Hamiltonian is nonintegrable owing to the 3PN spin-spin interaction resulting in the loss of the fifth integral.
Naturally, its equivalent Lagrangian with the accelerations up to the 2.5PN order is nonintegrable and can be chaotic.
Chaos is also possible in the Lagrangian with the accelerations up to the 1.5PN order.

\appendix
\section{Hamiltonian spin precession equations}

As in Sect. IIA, we still take
$\mathbf{S}_i=\mathbf{B}_i\times \mathbf{C}_i$, where $\mathbf{B}_i$ and $\mathbf{C}_i$
are regarded as \emph{generalized coordinates} and \emph{momenta}, respectively.
Unlike the  Lagrangian (6), the Hamiltonian (36) has no way to include the spin kinetic energy
but is slightly modified only in the expressional form
\begin{eqnarray}
H_1 &=& \frac{\mathbf{P}^{2}}{2\mu}-\frac{GM\mu}{R}+\frac{G}{c^{3} R^{3}}
(\mathbf{R}\times \mathbf{P})\cdot(\gamma_1\mathbf{B}_1\times \mathbf{C}_1 \nonumber \\
&& +\gamma_2\mathbf{B}_2\times \mathbf{C}_2).
\end{eqnarray}
The Hamilton's canonical equations with respect to the spin variables
are written as
\begin{eqnarray}
\mathbf{\dot{B}}_i &=& +\frac{\partial H_1}{\partial \mathbf{C}_i}=\frac{G\gamma_i}{c^{2} R^{3}}
(\mathbf{R}\times \mathbf{P})\times\mathbf{B}_i,\\
\mathbf{\dot{C}}_i &=& -\frac{\partial H_1}{\partial \mathbf{B}_i}=\frac{G\gamma_i}{c^{2} R^{3}}
(\mathbf{R}\times \mathbf{P})\times\mathbf{C}_i.
\end{eqnarray}
Then, we have the spin precession equations
\begin{eqnarray}
\mathbf{\dot{S}}_{iH_{1}} &=& \mathbf{\dot{B}}_i\times \mathbf{C}_i+\mathbf{B}_i\times \mathbf{\dot{C}}_i
 = \frac{G\gamma_i}{c^{2} R^{3}}
(\mathbf{R}\times \mathbf{P}) \times \mathbf{S}_i \nonumber \\
&=& \frac{\partial H_1}{\partial \mathbf{S}_i}\times \mathbf{S}_i.
\end{eqnarray}
It is worth pointing out that $\mathbf{B}_i$ and $\mathbf{C}_i$ do not mean the usual \emph{coordinates} and \emph{canonical momenta}
although they satisfy the  canonical equations (A2) and (A3).
They are completely different from the canonical, conjugate spin variables in [16], either.

\section{Evidences of chaos in the Lagrangian formalisms}

It was confirmed in [4] that there is chaos in
the simple dimensionless Lagrangian formalism of spinning compact binaries
with the Newtonian and leading-order spin-orbit contributions, i.e. $\mathcal{L}^{*}$ (23)
with the dimensionless acceleration $\mathbf{a}_1$ (17) for the spin transformations (16).
However, the right sides of the dimensionless spin equations (23) lose the factor $\mu/M$. This leads to minor errors.
Now, we reconsider the dynamics of correct descriptions of the dimensionless Lagrangian formalisms
$\mathcal{\widetilde{L}}^{*}$ (33) with $\mathbf{\widetilde{a}}_1$ (31) and $\mathcal{\widetilde{L}}^{*}$ with $\mathbf{\widetilde{a}}_2$ (32)
for the spin transformations (30). For comparison, the dynamics of $\widetilde{H}^{*}_1$ (56) is considered.

We take $c=G=1$ and the parameters $\beta=0.13$, $\chi_1=\chi_2=1$. The initial conditions
are $\mathbf{v}_{0}=(0,0.065,0)$, $\mathbf{r}_{0}=(24.5,0,0)$.
The two initial unit spin vectors are $\textbf{s}_1/|\textbf{s}_1|=(0.1,0.3,0.8)/\sqrt{0.1^{2}+0.3^{2}+0.8^{2}}$,
$\textbf{s}_2/|\textbf{s}_2|=(0.7,0.3,0.1)/\sqrt{0.7^{2}+0.3^{2}+0.1^{2}}$.
An eighth- and ninth-order Runge-Kutta-Fehlberg algorithm of variable step sizes [RKF8(9)]
is applied to work out the systems $\widetilde{H}^{*}_1$, $\mathcal{\widetilde{L}}^{*}$ with $\mathbf{\widetilde{a}}_1$ and  $\mathcal{\widetilde{L}}^{*}$ with $\mathbf{\widetilde{a}}_2$. This integrator can give high enough accuracies to the energy (21) (equivalently, the Hamiltonian (57)) or the Hamiltonian (56)
and therefore its numerical results should be reliable. Lyapunov exponents (see e.g. [17]), which
measure the average exponential deviation of two nearby orbits,  are suitable for quantifying  the ordered or
chaotic nature of  a dynamical system with any dimension. Unfortunately, sufficient long integration times
are generally necessary to make the values of Lyapunov exponents remain stable [18,19]. Instead,
a fast Lyapunov indicator (FLI) of two-nearby orbits [20] is regarded as a more sensitive tool to distinguish between chaos and order.
It is defined as
\begin{equation}
FLI=\log_{10}\frac{d_1}{d_0},
\end{equation}
where $d_0$ and $d_1$ denote the distances between two nearby orbits at times 0 and $t$, respectively.
The related details of this indicator were described in [20].

It is shown in Fig. 1 that the FLI for the 1.5PN Hamiltonian $\widetilde{H}^{*}_1$ grows slowly in a power law with time $\log_{10}t$.
This belongs to the characteristic of regularity. This result is expected because $\widetilde{H}^{*}_1$ is integrable.
However, the Lagrangian formalism $\mathcal{\widetilde{L}}^{*}$ with $\mathbf{\widetilde{a}}_2$ whose FLI increases exponentially
is chaotic due to the nonintegrability of its equivalent Hamiltonian $\widetilde{H}^{*}_2$.
Additionally, chaos exists in the Lagrangian formalism $\mathcal{\widetilde{L}}^{*}$ with $\mathbf{\widetilde{a}}_1$.
The chaoticity of the 1.5PN $\mathcal{\widetilde{L}}^{*}$ with $\mathbf{\widetilde{a}}_1$
and the regularity of the 1.5PN Hamiltonian $\widetilde{H}^{*}_1$ are consistent with the results of [4].

\section*{Acknowledgments}

This research has been supported by the National
Natural Science Foundation of China under Grant No. 11533004 and the Natural Science
Foundation of Jiangxi Province under Grant No. 20153BCB22001.

\begin{figure*}
\center{
\includegraphics[scale=0.3]{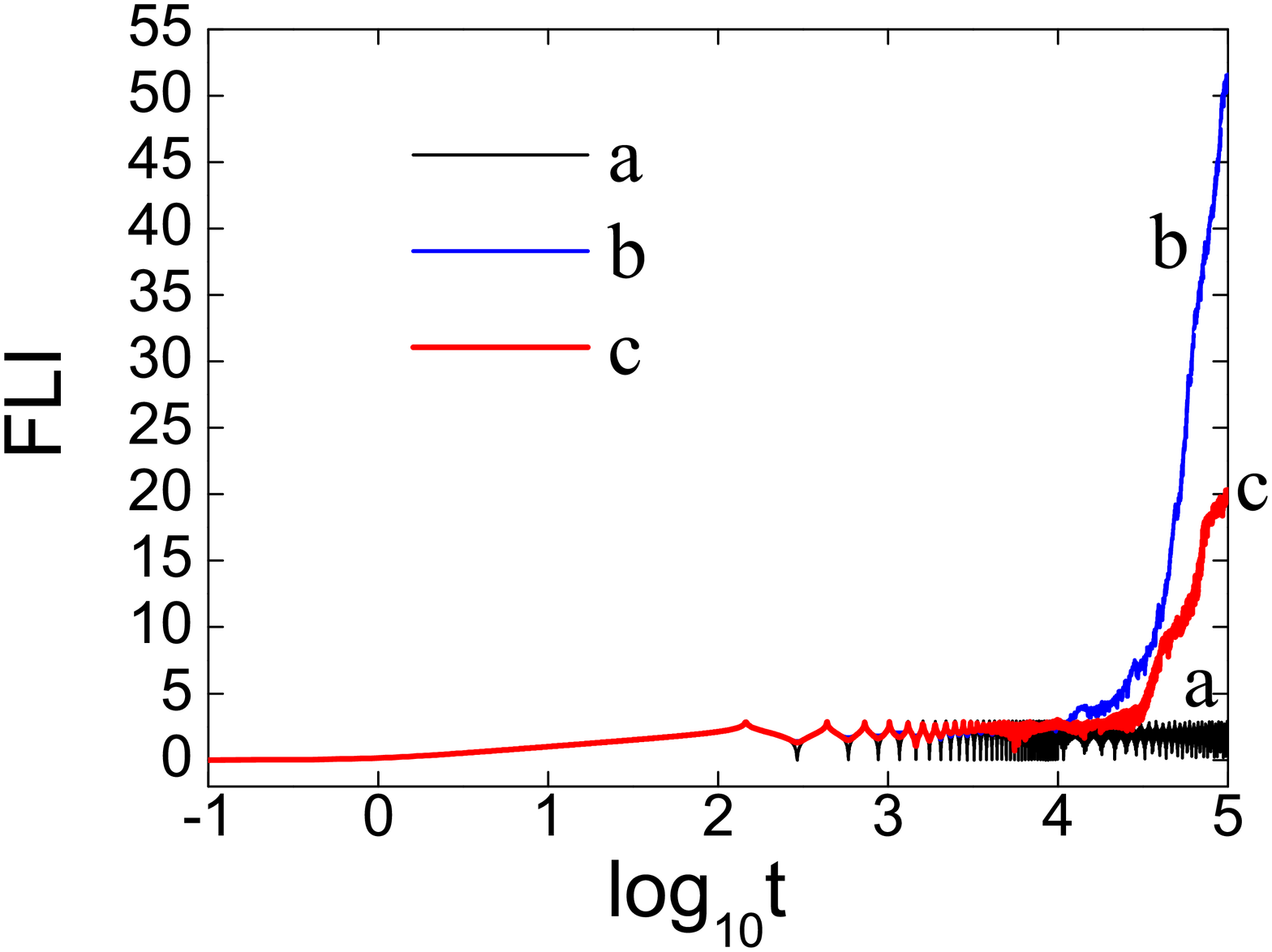}
\caption{Fast Lyapunov indicators (FLIs) for the three PN formalisms in the case of the scaling spin transformations $\mathbf{S}_i=\mathbf{s}_iG\mu M$.
(a) The 1.5PN Hamiltonian $\widetilde{H}^{*}_1$, (b) the Lagrangian formalism
$\mathcal{\widetilde{L}}^{*}$ with $\mathbf{\widetilde{a}}_1$ at 1.5PN order, and (c) the Lagrangian formalism $\mathcal{\widetilde{L}}^{*}$ with $\mathbf{\widetilde{a}}_2$
at 2.5PN order. $\widetilde{H}^{*}_1$ is typically regular, whereas $\mathcal{\widetilde{L}}^{*}$ with $\mathbf{\widetilde{a}}_1$ and  $\mathcal{\widetilde{L}}^{*}$ with $\mathbf{\widetilde{a}}_2$ are chaotic.
}} \label{fig1}
\end{figure*}

\end{document}